\documentclass[12pt,letterpaper]{article}
\usepackage[dvips]{graphicx,color}
\usepackage{amsmath,amssymb}
\usepackage[dvips,letterpaper,text={6.5in,9in}]{geometry}
\usepackage{fancyhdr}
\usepackage{verbatim}

\newcommand\ltap{\
  \raise.3ex\hbox{$<$\kern-.75em\lower1ex\hbox{$\sim$}}\ }
\newcommand\gtap{\
  \raise.3ex\hbox{$>$\kern-.75em\lower1ex\hbox{$\sim$}}\ }

\newcommand\simge{\mathrel{%
   \rlap{\raise 0.511ex \hbox{$>$}}{\lower 0.511ex \hbox{$\sim$}}}}
\newcommand\simle{\mathrel{
   \rlap{\raise 0.511ex \hbox{$<$}}{\lower 0.511ex \hbox{$\sim$}}}}

\newcommand{\slashchar}[1]%
        {\kern .25em\raise.18ex\hbox{$/$}\kern-.75em #1}
\def\lsim{\mathrel{\raise.3ex\hbox{$<$\kern-.75em\lower1ex\hbox{$\sim$}}}}
\def\gsim{\mathrel{\raise.3ex\hbox{$>$\kern-.75em\lower1ex\hbox{$\sim$}}}}
\newcommand{\bs}{\boldsymbol}

\newcommand\CH{{\cal H}}

\newcommand\CL{{\cal L}}

\newcommand\CO{{\cal O}}

\newcommand\be{\begin{equation}}
\newcommand\ee{\end{equation}}
\newcommand\bea{\begin{eqnarray}}
\newcommand\eea{\end{eqnarray}}
\newcommand\ba{\begin{array}}
\newcommand\ea{\end{array}}
\newcommand\nn{\nonumber}

\newcommand{\half}{\ensuremath{\frac{1}{2}}}

\newcommand\ra{\rightarrow}

\newcommand\ol{\bar}
\newcommand\mev{{\rm MeV}}
\newcommand\gev{{\rm GeV}}
\newcommand\tev{{\rm TeV}}

\newcommand\lvac{\langle \Omega \vert}
\newcommand\rvac{\vert \Omega \rangle}

\newcommand\uy{U(1)_Y}
\newcommand\sutwow{SU(2)_W}
\newcommand\sutwop{SU(2)'}
\newcommand\sutwoc{SU(2)_C}

\begin{document}
\title{
\vskip -15mm
\begin{flushright}
\vskip -15mm
{\small BUHEP-05-17\\
hep-ph/0511002\\}
\vskip 5mm
\end{flushright}
{\Large{\bf A New Mechanism for Light Composite Higgs Bosons}}\\
}
\author{
{\large Kenneth Lane\thanks{lane@bu.edu}\,\, and Adam
    Martin\thanks{aomartin@bu.edu}}\\
{\large {$$}Department of Physics, Boston University}\\
{\large 590 Commonwealth Avenue, Boston, Massachusetts 02215}\\
}
\maketitle

\begin{abstract}
  Repeated symmetry-breaking and restoration phase transitions occur as one
  traverses the parameter space of interactions competing to align the
  vacuum. This phenomenon, augmented with a topcolor-like interaction, can make
  a composite Higgs boson's mass and vacuum expectation value naturally
  much less than its underlying structure scale, without introducing new
  symmetries and their associated TeV-scale particles. We illustrate it by
  reconstructing a simple light composite Higgs model of electroweak symmetry
  breaking proposed by Georgi and Kaplan.
\end{abstract}


\newpage


Precision data strongly suggest that spontaneous electroweak
$\sutwow\otimes\uy$ symmetry breaking (EWSB) occurs via a scalar doublet with
low Higgs boson mass $M_h \simle 300\,\gev$~\cite{Group:2005em}. On the other
hand, EWSB via an elementary Higgs boson field, without new physics to
stabilize its mass and VEV at such a low energy, is fraught with well-known
theoretical difficulties. There are many interesting proposals for
stabilizing the Higgs mass, yet none has led to a completely satisfactory
``standard model of physics beyond the standard model''. Therefore, it is
worthwhile to explore many approaches; one of them may even be supported by
data from the LHC. We propose here a new variant of light composite Higgs
(LCH) models, one in which the Higgs is an accidental Goldstone boson
(AGB)~\cite{Lane:2005we}.

An AGB is a pseudo-Goldstone bosons which is anomalously light compared to
its PGB partners because its mass must vanish at second-order symmetry-phase
transitions and because such transitions can occur repeatedly as one varies
one or more couplings in the explicit chiral symmetry breaking interaction
$\CH'$ that gives mass to the PGBs. The phase transitions occur on critical
surfaces in $\CH'$-parameter space. Varying one coupling follows a trajectory
in this space and a transition occurs as it pierces a critical surface. If we
happen to live in a region between two transitions, at least one PGB is likely
to be very light. In Ref.~\cite{Lane:2005we} we studied this phenomenon,
emphasizing that the generality of phase transitions and the exact vanishing
of the AGB mass there. There we focused on discrete symmetry transitions but,
of course, the same occurs for continuous symmetries, and that will be
important for us.

In~\cite{Lane:2005we} we considered AGBs with a specific ultraviolet
completion: They were $\ol \psi_L \psi_R$ composites of $N$ strongly
interacting massless Dirac fermions $\psi_i$ whose characteristic interaction
scale is $\Lambda_\psi \simeq 4\pi F_\pi$, with $F_\pi$ the PGBs' decay
constant. We suggested there that an AGB may serve as an LCH for EWSB. To
qualify as an LCH, the boson must be much lighter than $\Lambda_\psi$ {\em
  and} its VEV $v$ much less than
$F_\pi$~\cite{Kaplan:1983fs,Kaplan:1983sm,Georgi:1984ef,Georgi:1984af,
  Dugan:1984hq}. Following common practice, we will take $v \simeq250\,\gev
\ll F_\pi$ to mean $v < (0.25$--$0.5) F_\pi$, postponing the
naturalness/hierarchy problem of standard elementary Higgs models to
$\Lambda_\psi = 5$--$10\,\tev$.

Light AGB masses, due to multiple phase transitions, were easy to achieve
in~\cite{Lane:2005we}. While we also showed that $v \ll F_\pi$ may occur, we
gave no recipe for it. The problem was that those were {\em discrete}
symmetry transitions and the VEV did not always need to vanish when the mass
did. Transitions between phases of a continuous symmetry correlate the
vanishing of the mass and the VEV, and that is the basis of our approach to
constructing LCH models of EWSB: A set of interactions in $\CH'$, some of
which come from physics well above $\Lambda_\psi$, compete to align the
vacuum in different directions, creating a complex vacuum structure. As one
follows paths in the space of $\CH'$ couplings, one encounters multiple
spontaneous breaking and restorations of electroweak symmetry. This drives
the Higgs mass and VEV repeatedly to zero, with the possibility that {\em
  both} remain small in the intervening EWSB region we live in.

This AGB mechanism is natural in that $M_h$ and $v$ can stay small for a
sizable region, $\Delta\kappa$, of a basic coupling $\kappa$ in $\CH'$.  We
emphasize, however, that there is no symmetry in our scheme keeping $M_h$ and
$v$ small. If the spacing $\Delta\kappa$ between phase transitions is too
large, $v$ grows to remain close to $F_\pi$ over most of the region. If
$\Delta\kappa/\kappa \ll 1$, $M_h$ and $v$ would be tiny, but at the expense
of fine-tuning. We find that it is easy to choose parameters so that both
remain small over a region with $\Delta\kappa/\kappa = \CO(1)$. We shall
quantify this statement below. An advantage of not imposing additional
symmetries is that we need not introduce associated TeV-scale particles that
can conflict with precision measurements.

In this paper we illustrate the AGB mechanism with a toy model, albeit a
fairly sophisticated one. It is based on Refs.~\cite{Georgi:1984af,
  Dugan:1984hq}. It has the nice feature of a custodial $\sutwoc$ symmetry
preserving $\rho \cong 1$. The model's additional scalars are weakly coupled
to ordinary matter and, so, have little impact on other precisely measured EW
quantities. Given the large mass of the top quark, we will have to invoke
topcolor-like gauge interactions~\cite{Hill:1994hp} to prevent its
overwhelming influence on vacuum alignment. This is a general feature of our
scheme. Taking the particular model seriously, it has some interesting
phenomenology that we'll discuss at the end.

Note the differences between our scheme and that of little Higgs
models~\cite{Arkani-Hamed:2001nc,Arkani-Hamed:2002qy,
  Arkani-Hamed:2002qx,Schmaltz:2002wx}. There, to ensure that one is close to
the EW transition without fine tuning once quantum corrections are included,
approximate global symmetries, involving new heavy particles, are imposed to
soften the cutoff dependence. Furthermore, the top quark plays a central role
in breaking EW symmetry in little Higgs models. In topcolor, the contribution
of the $t \bar t$ condensate to $v$ is small.


{\em The Model:} To cast our proposal in familiar terms, the model we use is
based on an $SU(5)/SO(5)$ symmetry breaking pattern, as are the models of
Refs.~\cite{Georgi:1984af,Dugan:1984hq,Arkani-Hamed:2002qy}. In fact, we
follow the pattern of the Georgi-Kaplan (GK) model exactly, relying
extensively on its full description in Ref.~\cite{Dugan:1984hq}.

The model has five flavors of massless Weyl fermions $\psi = \{\psi_i,\,
i=1,\dots,5\}$ transforming as the real representation of some strong
``ultracolor'' group, $G_{UC}$. It is assumed that $G_{UC}$ interactions form
condensates, given in the ``standard vacuum'' $\rvac$ by
\bea\label{eq:Tcond}
\lvac \psi_i \psi^T_j \rvac &\simeq& - 2\pi F_\pi^3\Delta_{ij}\,, \\
\Delta &=& \left(\ba{cc} \sigma_2 \otimes \tau_2 & 0\\
                         0 & 1\\ \ea\right)\,.
\eea
The vacuum symmetry $SO(5)$ contains $\sutwow \otimes \sutwop$ --- the gauged
electroweak $SU(2)$ symmetry (coupling $g$) and an $SU(2)$ whose third
generator is weak hypercharge $\uy$ ($Y = Q'_3$ with coupling $g'$).
\be\label{eq:generators}
Q_a = \half \left(\ba{cc} \sigma_a \otimes 1 & 0\\
                            0 & 0 \ea\right) \,,\qquad
Q'_a = \half \left(\ba{cc} 1 \otimes \tau_a & 0\\
                            0 & 0 \ea\right)\,;\,\,\,(a=1,2,3)\,.
\ee
When EWSB occurs, $\sutwow \otimes \sutwop \ra \sutwoc$, the custodial
$SU(2)$. Generators $T_a$ of $SO(5)$ and $X_a$ of $SU(5)/SO(5)$ satisfy
$\Delta T_a \Delta = - T^T_a$ and $\Delta X_a \Delta = X^T_a$. An important
generator is $Q_{24} = 1/\sqrt{20} \, {\rm diag}(1,1,1,1,-4)$. We will assume
it is gauged, but broken far above $\Lambda_\psi$.

The fluctuations of the vacuum about the standard one with condensate
$\Delta$ are described by the unitary matrix $U$:
\be\label{eq:Umatrix}
U = e^{i {\bs{H}}/2F_\pi}
e^{i{\bs{\eta}}/F_\pi} e^{i{\bs{\pi}}/F_\pi}
e^{i{\bs{H}}/2F_\pi} \Delta \equiv \Sigma \Delta\,,
\ee
where the 14 PGB fields and their $\sutwow\otimes\sutwop$ representations are
\bea\label{eq:PGBs}
 {\bs{H}} &=& \frac{1}{\sqrt{2}} \left(\begin{array}{ccc} & & \tilde h \\ & &
     h \\ \tilde h^{\dag} & h^{\dag} & 0  \\ \end{array} \right) \,\,\in\,\,
 (2,2) \,, \qquad  h =  \left(\begin{array}{c} h_1 + i~h_2 \\ h_0 + i~h_3
     \\ \end{array}  \right)\,,\,\,
 \tilde h = i\sigma_2 h^*\,; \nn \\ \\
{\bs{\eta}} &=& \sqrt{2}\, \eta\, Q_{24} \,\,\in\,\, (1,1)\,; \qquad
{\bs{\pi}} = \frac{1}{\sqrt{2}}\, \pi_{ab}
                    \left(\ba{cc} \sigma_a \otimes \tau_b & 0\\
                                   0 & 0\ea\right) \,\,\in\,\, (3,3)\,.\nn
\eea
When EWSB occurs and the three GBs in $h$ are eaten, the Higgs field is given
in unitary gauge by ${\bs {H}} = hX/F_\pi$ where $X_{ij} = (\delta_{i1} +
\delta_{i4})\delta_{j5} + (i \leftrightarrow j)$. So long as only the Higgs
doublet gets a VEV, $v = \langle h\rangle$, the expectation value of $\Sigma$
is $\langle\Sigma\rangle = e^{i\langle {\bs{H}}\rangle} = 1 + i \sin(v/F_\pi)
X + (\cos(v/F_\pi) - 1) X^2$ and the weak boson masses are, to
$\CO(g^2\alpha)$,
\be\label{eq:WZmasses}
M_W^2 = M_Z^2 \cos^2\theta_W = \half g^2 F_\pi^2 (1-\cos(v/F_\pi))\,.
\ee
%


{\em Symmetry Breaking Interaction $\CH'$:} The chiral symmetries of $\Sigma$
are explicitly broken by a Hamiltonian $\CH'$ which receives contributions of
$\CO(g^2,g^{'2})$ from the electroweak interactions. Additional contributions
come from broken ``extended ultracolor'' (EUC) interactions of the
$\psi$-fermions. They are mediated by heavy gauge bosons with typical mass
$M_E \gg \Lambda_\psi$ and coupling $\alpha_E$. We assume this $G_{UC}$
``walks''~\cite{Holdom:1981rm,Appelquist:1986an,Akiba:1986rr,Yamawaki:1986zg}.
Then, the usual $g_E^2/M_E^2$ suppression factor is enhanced by $\approx
(M_E/\chi F_\pi)^{2\gamma_\psi}$, where $\gamma_\psi$ is the $\psi\psi^T$
anomalous dimension, equal to one in a strictly walking gauge theory and
otherwise somewhat less, and $\chi \simeq 1$ to $4\pi$, depending on where
the anomalous dimension integral runs from.

In general, one expects several relevant EUC operators contributing to
$\CH'$, much as in the model we used to study the AGB mechanism in
Ref.~\cite{Lane:2005we}. The resulting vacuum structure is then quite
complex. To illustrate our AGB mechanism simply, we assume just one EUC
operator competes with the electroweak ones, and that $\CH'$ is
$\sutwow \otimes \sutwop$-invariant:
\bea\label{eq:Hprime}
\CH' &=& -F_\pi^4 \bigl[c_1{\rm Tr}\, (Y \Sigma Y\Sigma^{\dag}) 
          + c_2 \sum_{a=1}^3 {\rm Tr}\, (Q_a \Sigma Q_a
          \Sigma^{\dag}) \nn\\
       && \quad\,\,\,\, + c_3{\rm Tr}\, (Q_{24} \Sigma Q_{24} \Sigma^\dagger)
        + c_4 \sum_{a=1}^2 {\rm Tr}\, (P_a \Sigma P_a^\dagger
       \Sigma^\dagger + P_a^\dagger \Sigma P_a \Sigma^\dagger)\bigr]\,.
\eea
As noted above, the top mass arises from topcolor, so that $\Sigma$'s
coupling to top is small. The first two terms in $\CH'$ are in the GK model,
with $c_{1} = \CO(g'^2)$, $c_2 = \CO(g^2)$ and positive. In GK, a gauge boson
coupled to $Q_{24}$ eats the $\eta$ and produces the third term with $c_3 <
0$. It then competes with the first two terms over the fate of electroweak
symmetry. There is a single single phase transition at a critical value of
$c_3$. Then, the range of $c_3$ for which $M_h^2$, $v^2 \ll F_\pi^2$ is quite
small, so that it must be tuned to obtain an LCH. Taken seriously, moreover,
this origin for $c_3$ makes it approximately $4\pi^2$. Assuming it arises
from EUC interactions can make it smaller. The magnitude and sign of $c_3$ in
our scenario are discussed below. The $c_4$ term, with $(P_a)_{ij} =
\delta_{ia} \delta_{j5}$, is included solely to give mass to the $\eta$. Our
mechanism is most simple if $c_4$ is small. On the other hand, vacuum
alignment is numerically difficult when massless particles are present.
Therefore, we take $c_4$ small and positive.\footnote{The consequences of a
  very light $\eta$ will be discussed later.}

With EW symmetry intact ($v=0$), $\CH'$ gives the following masses to the
four degenerate Higgs particles $h$, six $\pi$ and three $\pi'$ from the
(3,3) (split by $\uy$), and the $\eta$:
\bea\label{eq:PImasses}
M^2_h &=& \textstyle{\frac{1}{2}}(c_1 + 3 c_2 + 5 c_3 + 10 c_4)
F_\pi^2 \nn \\ 
M^2_{\pi} &=& 2(c_1 + 2 c_2 + c_4)F_\pi^2 \nn \\
M^2_{\pi'} &=& 2(2 c_2 + c_4)F_\pi^2 \nn \\
M^2_{\eta} &=& 10 c_4 F_\pi^2 \,.
\eea

For the $Q_{24}$ term in $\CH'$, we regard the EUC coupling $\alpha_E$, not
$c_3$, as the fundamental parameter. Because ultracolor walks, $\alpha_E
\simeq \alpha_{UC}$ and it is slightly less than the critical coupling for
$\psi\psi^T$ condensation, approximately $\pi/3 C_2(R_\psi)$. The quadratic
Casimir $C_2(R_\psi) \simeq N_{UC}$, the number of ultracolors. The
$\CO(\alpha_E^2)$ contribution to $c_3$, an ultracolor radiative correction,
is about $3\alpha_E N_{UC}/4\pi$ times the $\CO(\alpha_E)$ contribution and
that is not negligible.\footnote{Appelquist, Bai and
  Piai~\cite{Appelquist:2005iz} recently used higher-order operators to tilt
  the vacuum in symmetry breaking directions. We differ from them in our
  simplifying assertion that higher-order corrections to a single operator
  can induce repeated symmetry breaking.} Therefore, with $\kappa =
4\pi\alpha_E \simeq 1$--10, we write
\bea\label{eq:Cfactors}
c_3 &=& a_3 \kappa + b_3\kappa^2\,;\nn\\
|a_3| &\simeq& \frac{1}{4 M^2_E} \Lambda_\psi^2
\left(\frac{M_E}{\chi F_\pi}\right)^{2\gamma_\psi} = 
\frac{1}{4}\left(\frac{4\pi}{\chi}\right)^{2\gamma_\psi} 
\left(\frac{\Lambda_\psi^2}{M^2_E}\right)^{1-\gamma_\psi} \\
\left|\frac{b_3}{a_3}\right| &\simeq& \frac{3N_{UC}}{16\pi^2} = \CO(0.1)
\quad {\rm for}\,\, N_{UC} \simeq 4\,. \nn
\eea
The factor $|a_3| = \frac{1}{4}(4\pi/\chi)^{2\gamma_\psi}
(\Lambda_\psi^2/M_E^2)^{1-\gamma_\psi}$ is sensitive to its parameters and
can easily lie in the range 0.1--10.

For an interaction mediated by heavy gauge boson exchange, it is entirely
plausible that $a_3 < 0$, as GK assumed. The coefficient $b_3$ may have
either sign; we assume $b_3 > 0$. We choose $c_4$ to be a small positive
constant so that $M_\eta^2 > 0$. Let $2.5c_3 = A\kappa + B\kappa^2$ and $C =
0.5(c_1+3c_2 + 10c_4)$. Then, if $A^2 - 4BC > 0$ --- an inequality we expect,
given the origin of these terms --- there are two critical values of
$\kappa$, $\kappa_\mp^* = (-A \mp \sqrt{A^2 - 4BC})/2B$, at which $M_h^2$
vanishes and EW symmetry is broken and then restored. The relative size of
the EWSB region is
\be\label{eq:delkappa}
\frac{\Delta\kappa}{\bar\kappa} \equiv \frac{\kappa_+^* - \kappa_-^*}
{(\kappa_+^* + \kappa_-^*)/2} =  2\sqrt{1 - 4BC/A^2}\,.
\ee
This provides one measure of parameter tuning in our scheme. We shall require
that $\Delta\kappa/\bar\kappa \simeq 0.5$--1, i.e., no fine-tuning of the
coupling $\kappa$.

Another measure of tuning is the amount $A$ can be varied (for fixed $B/A$)
while maintaining $\Delta\kappa/\bar\kappa \simeq 0.5$--1. For nominal values
of $c_1$ and $c_2$, $C \simeq 0.5$. Then, with $|B/A| = \CO(0.1)$, $|A|$
should be comparable to $C$ so that $A\kappa + B\kappa^2 + C$ can vanish
twice in the region $\kappa \simeq 1$--10. As just noted, this is a
reasonable estimate of $A$ (and $B$). In the example described below, we find
that we can vary $A$ by $\pm 10\%$ (i.e., $\Delta A/\bar A \simeq 0.2$) about
a central value and still have $\Delta\kappa/\bar\kappa \simeq 0.5$--1 and
$M_h^2/F_\pi^2$, $v^2/F_\pi^2 \ll 1$. For small $v^2/F_\pi^2$, the Higgs
self-interaction in $\CH'$ is well-approximated by a quartic potential, so
that the Higgs mass satisfies
\be\label{eq:hmass}
M_h^2 \cong 2\lambda_h v^2 \simle (M_h^2)_{\rm max} = 2(A^2/4B -C)F_\pi^2\,,
\ee
where the Higgs quartic self-coupling is $\lambda_h = -(c_1 + 3c_2 +
20c_3 + 22c_4)/12$.

 \begin{figure}[t]
   \begin{center}
     \includegraphics[width=2.75in,height=3.10in, angle=270]{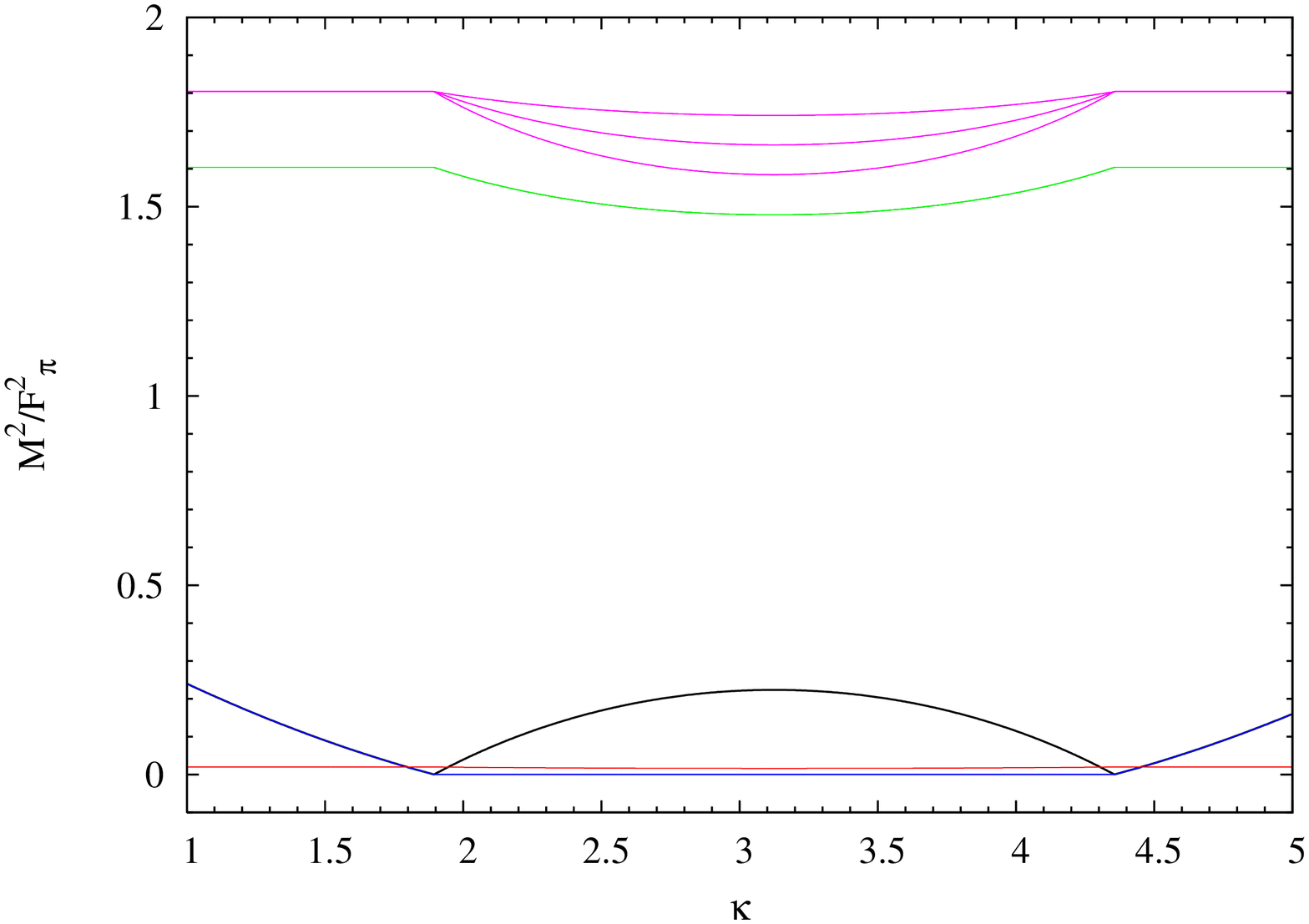}
     \includegraphics[width=2.75in,height=3.10in, angle=270]{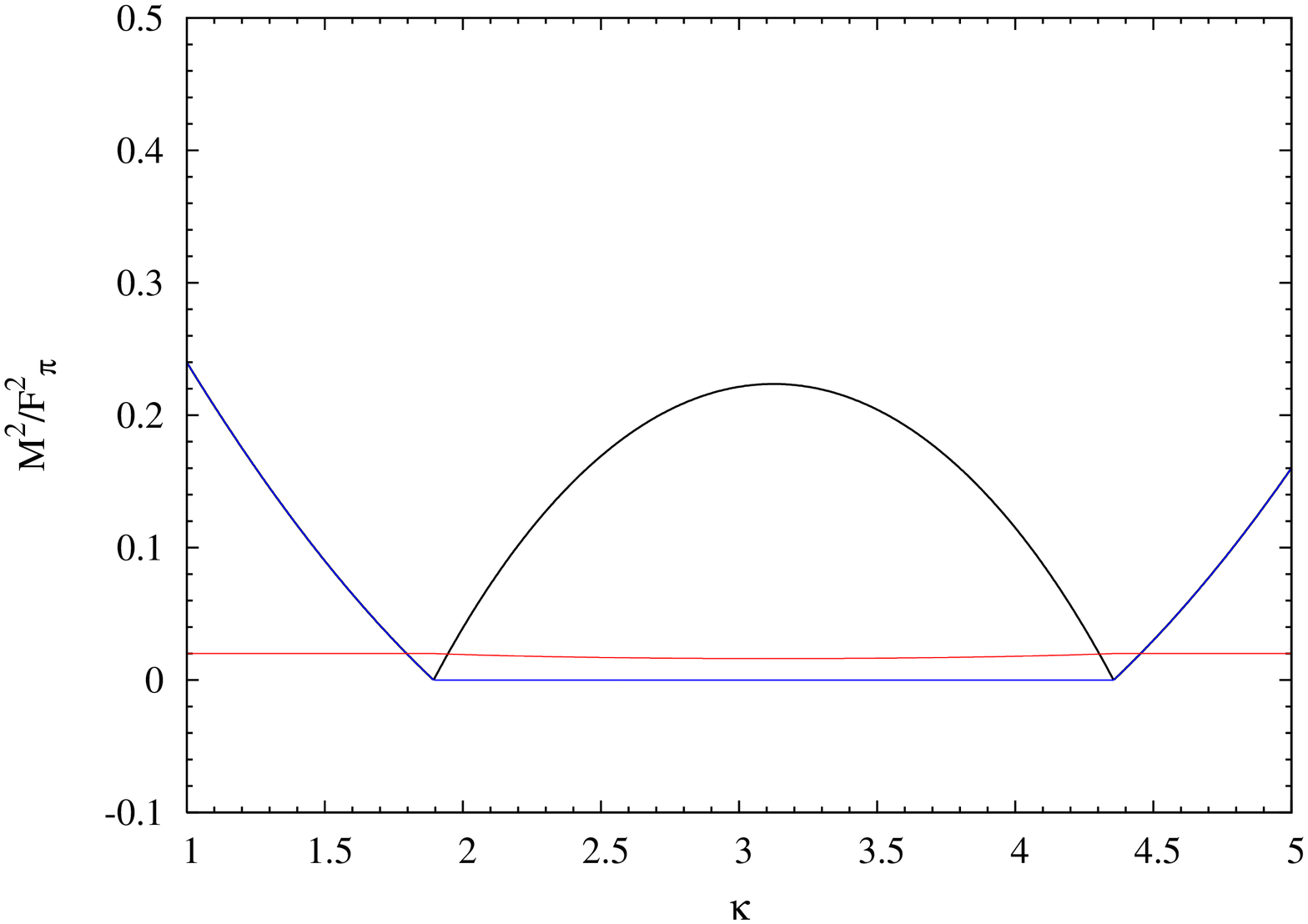}
     \caption{Squared masses relative to $F_\pi^2$ of the six $\pi$
       (magenta), three $\pi'$ (green), $\eta$ (red) and four Higgs bosons
       (black and blue) on the left, and a closeup of the Higgs masses in the
       EWSB region (right) in the $SU(5)/SO(5)$ model with parameters stated
       in the text.}
     \label{fig:AGB_EWSB_fig_1-2}
   \end{center}
 \end{figure}
\begin{figure}[t]
   \begin{center}
     \includegraphics[width=2.75in,height=3.10in, angle=270]{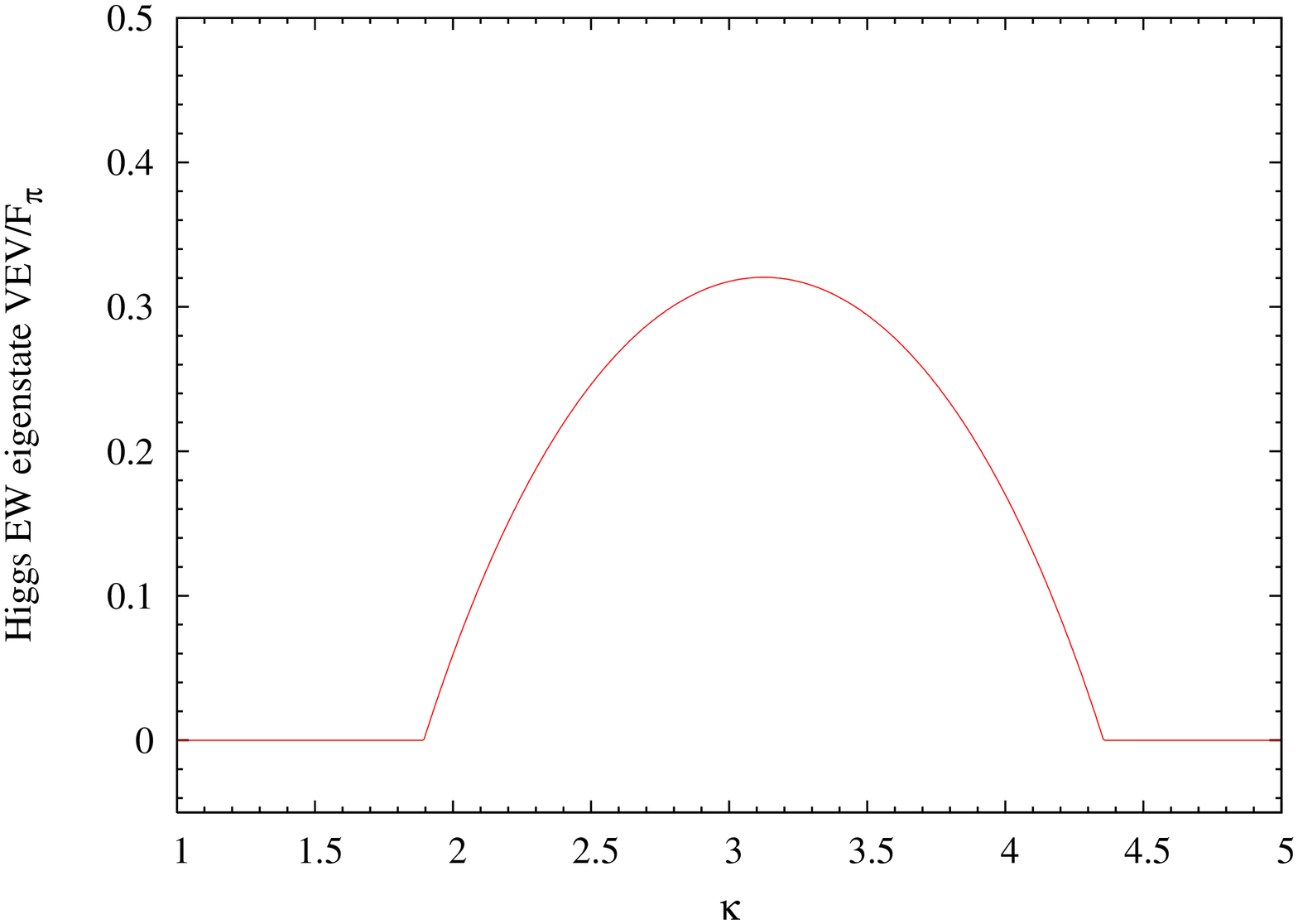}
     \includegraphics[width=2.75in,height=3.10in, angle=270]{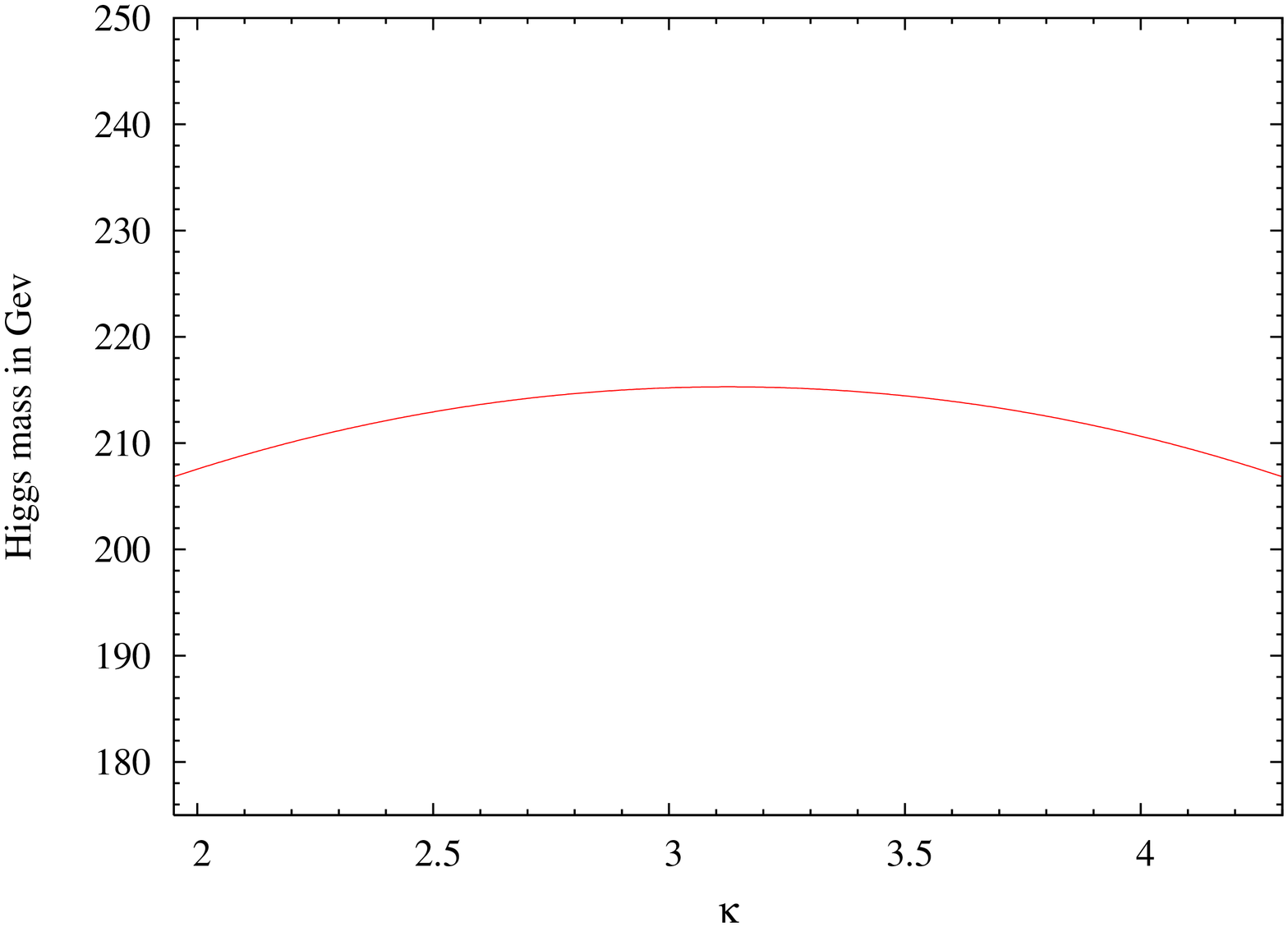}
     \caption{The Higgs boson VEV $v^2/F_\pi^2$ (left) and $M_h(\gev)$ for $v =
     246\,\gev$ in the $SU(5)/SO(5)$ model with parameters stated in the
     text.}
     \label{fig:AGB_EWSB_fig_3-4}
   \end{center}
 \end{figure}

 A specific example is provided by $c_1 = 0.1$, $c_2 = 0.4$, $c_3 =
 0.2(-\kappa + 0.16 \kappa^2)$, and $c_4 = 0.002$. Vacuum alignment is
 carried out numerically as described in
 Refs.~\cite{Lane:2005we,Lane:2000es}. The PGB masses are shown in
 Fig.~\ref{fig:AGB_EWSB_fig_1-2}. Varying $\kappa$ from one to five, only $h$
 gets a nonzero VEV, which occurs in the region $1.89 \simle \kappa \simle
 4.36$.  In the EWSB region, note the massless (eaten) Goldstone bosons and
 the splitting of the charged and neutral members of the $\pi$-sextet. The
 Higgs VEV $v^2/F_\pi^2$ is shown on the left
 Fig.~\ref{fig:AGB_EWSB_fig_3-4}. In this example, $v \simle 0.5 F_\pi$ or,
 fixing $v = 246\,\gev$, $\Lambda_\psi \simge 5\,\tev$. We plot the Higgs
 mass, with $v$ fixed, on the right in Fig.~\ref{fig:AGB_EWSB_fig_3-4},
 obtaining $M_h \simeq 215\,\gev$. Then, $M_{\pi,\pi'} \simeq 600\,\gev$
 while the $\eta$ is very light. Varying $A$ by $\pm 10\%$,
 $\Delta\kappa/\bar \kappa$ changes from~0.9 to~0.5 and $F_\pi$ from 400 to
 700~GeV, but $(M_h)_{\rm max}$ only changes from 230 to 210~GeV --- because
 $\lambda_h$ varies slightly.

 The new degrees of freedom below 1~TeV in our toy model are the nine $\pi_a$
 and the $\eta$. That there are no new $W'$ nor heavy quarks as in
 little Higgs models is an attractive model-independent feature of our
 scheme. The new scalars are weakly-coupled to the $W$ and $Z$, quarks and
 leptons, and to each other. The GK model has a parity-like symmetry that
 requires them to be emitted and/or absorbed in pairs. Thus, they have little
 effect on precisely-measured quantities such as the $S$-parameter,
 forward-backward asymmetries, etc.
 
 A more model-dependent feature is the $\eta$-mass. We made the $\eta$ very
 light to avoid complicating $\CH'$ with extra operators. Amusingly enough,
 we believe $\eta$ could be practically massless and, with $F_\pi \simge
 5\,\tev$, still have evaded the searches and tests for an axion because it
 must be pair-produced. Furthermore, the $h^2\eta^2$ coupling
 $\lambda_{h\eta} \cong -5(c_1 + 3c_2 + 5c_3 + 10c_4)/32$ implies $\Gamma(h
 \ra \eta\eta) = 4.0\,\mev \simeq 0.8\Gamma(h \ra \bar b b)$, making the
 Higgs of this model somewhat harder to find than the standard-model one.
 
 The masses of quarks and leptons --- except for the top --- arise
 technicolor-style from their EUC couplings to $\psi$-fermions (also see
 Refs.~\cite{Georgi:1984af,Dugan:1984hq}). To lowest order in $g_E^2/M_E^2$,
 these interactions produce
\be\label{eq:Yukawa}
\CL_Y = \Gamma^d_{\alpha\beta}\, F_\pi\, q_{i\alpha} \,\Sigma_{ij} \,
d^c_{j\beta} + \cdots
\,,
\ee
where $q_{i\alpha} = u_\alpha\delta_{i1} + d_\alpha\delta_{i2}$,
$d^c_{i\alpha} = d^c_\alpha \delta_{i5}$. The Yukawa couplings $\Gamma^{q,l}$
are enhanced by the walking $G_{UC}$ interaction. These interactions are
naturally flavor conserving~\cite{Glashow:1976nt}.

We expect $\Gamma^t \approx \Gamma^b = m_b/v$ or perhaps somewhat larger,
while most of the top mass comes from $\bar tt$ condensation induced by
topcolor-like interactions. Because $\langle \bar tt \rangle$ contributes
little to EWSB~\cite{Hill:1994hp}, we ignored it in fixing $v = 246\,\gev$ to
make our mass estimates. Standard topcolor requires a color-octet $V_8$ and a
singlet $Z'$ to produce $m_t \gg m_b$. They are expected to have masses of
several~TeV.

To sum up, we have argued that light composite Higgs models may be
constructed without excessive tuning and without the need for extra particles
canceling large loop corrections to $M_h$. This happens when the Higgs is an
accidental Goldstone boson. That is, its mass and VEV are small because we
live in a region of $\CH'$ space that lies between successive EWSB phase
transitions --- at which $M_h$ and $v$ must vanish. A general feature of our
scheme is that a topcolor-like interaction is needed to minimize the top's
influence on EWSB. We illustrated our mechanism using a composite Higgs
model due to Georgi and Kaplan for which we proposed a plausible dynamical
origin for the operators in $\CH'$. We limited the number of these operators
to keep our exposition as simple as possible. But, as we have
argued~\cite{Lane:2005we}, we believe this AGB phenomenon is quite general,
brought on by the competition to align the vacuum among several terms in
$\CH'$. And, while the model is a toy, it is consistent with precision
measurements pointing to a standard electroweak model with a light Higgs
boson.

We are grateful to Howard Georgi and Martin Schmaltz for critical comments.
We also thank Nima Arkani-Hamed, Tom Appelquist, Fawzi Boudjema, Andy Cohen,
Eric Pilon and Pierre Sikivie for valuable discussions. KL thanks the
Laboratoire d'Annecy-le-Vieux de Physique Theorique, Annecy, France for its
hospitality and partial support for this research during the summer of 2005.
He also thanks the Aspen Center for Physics for its stimulating atmosphere.
This research was also supported by the U.~S.~Department of Energy under
Grant~No.~DE--FG02--91ER40676.


\bibliography{AGB_EWSB.bib}
\bibliographystyle{utcaps}
\end{document}